\documentclass[journal=jpccck,manuscript=article]{achemso} 
\usepackage{chemformula} 
\usepackage[T1]{fontenc} 
\usepackage[version=3]{mhchem} 
\usepackage{indentfirst}

\author{Mingda Ding}
\affiliation[OUKobayashi]
{Department of Applied Physics, Graduate School of Engineering, Osaka University, Suita, Osaka 565-0871, Japan}
\author{Taiki Inoue}
\affiliation[OUKobayashi]
{Department of Applied Physics, Graduate School of Engineering, Osaka University, Suita, Osaka 565-0871, Japan}
\author{John Isaac Enriquez}
\affiliation[OUMori]
{Department of Precision Engineering, Graduate School of Engineering, Osaka University, Suita, Osaka 565-0871, Japan}
\author{Harry Handoko Halim}
\affiliation[OUMori]
{Department of Precision Engineering, Graduate School of Engineering, Osaka University, Suita, Osaka 565-0871, Japan}
\author{Yui Ogawa}
\affiliation[NTT]
{NTT Basic Research Laboratories, NTT Corporation, Kanagawa 243-0198, Japan}
\author{Yoshitaka Taniyasu}
\affiliation[NTT]
{NTT Basic Research Laboratories, NTT Corporation, Kanagawa 243-0198, Japan}
\author{Yuji Hamamoto}
\affiliation[OUMori]
{Department of Precision Engineering, Graduate School of Engineering, Osaka University, Suita, Osaka 565-0871, Japan}
\author{Yoshitada Morikawa}
\affiliation[OUMori]
{Department of Precision Engineering, Graduate School of Engineering, Osaka University, Suita, Osaka 565-0871, Japan}

\author{Yoshihiro Kobayashi}
\affiliation[OUKobayashi]
{Department of Applied Physics, Graduate School of Engineering, Osaka University, Suita, Osaka 565-0871, Japan}
\email{kobayashi@ap.eng.osaka-u.ac.jp}
\phone{+81-06-6879-7833}
\fax{+81-06-6879-7863}

\title{Reduction of interlayer interaction in multilayer stacking graphene with carbon nanotube insertion: Insights from experiment and simulation}

\abbreviations{IR,NMR,UV}

\begin{document}

\begin{tocentry}

\
\newline

\includegraphics[keepaspectratio,width=8.2cm]{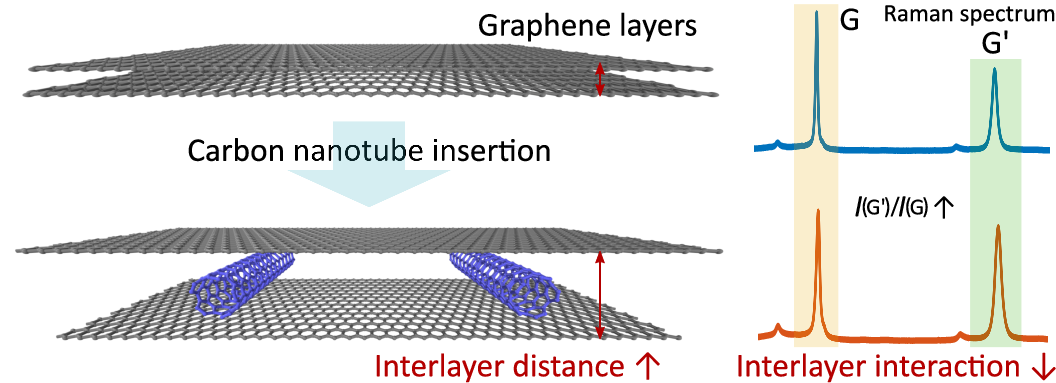}




\end{tocentry}

\begin{abstract}
The creation of multilayer graphene, while preserving the brilliant properties of monolayer graphene derived from its unique band structure, can expand the application field of graphene into macro scale. However, the energy-favorable AB stacking structure in the multilayer graphene induces a strong interlayer interaction and alters the band structure. Consequently, the intrinsic properties of each monolayer are degraded. In this work, we insert carbon nanotubes (CNT) as nanospacers to modulate the microstructure of multilayer stacking graphene. Nanospacers can increase interlayer distance and reduce interlayer interaction. The graphene/CNT stacking structure is experimentally fabricated using a dry transfer method in a layer-by-layer manner. Raman spectroscopy verifies the reduction in interlayer interaction within the stacking structure. Atomic force microscopy scanning shows an increase in the interlayer distance, which can explain the weakening of interlayer interactions. The microstructure of the stacked graphene and CNTs is studied by molecular dynamics simulation to systematically investigate the effect of CNT insertion. We found that the distribution distance, size, and arrangement of the CNT can modulate the interlayer distance. These results will help to understand and improve the properties of the composite systems consisting of graphene and CNTs.
\end{abstract}


\section{Introduction}
Graphene is a material with excellent electrical, thermal, and optical properties\cite{novoselov_two-dimensional_2005,meric_current_2008,mak_measurement_2008,balandin_superior_2008,lee_changgu_measurement_2008} and broad prospects for applications\cite{bunch_electromechanical_2007,koppens_photodetectors_2014,raccichini_role_2015,deng_wrinkled_2016,akinwande_review_2017}. Monolayer graphene has a unique band structure with a linear electronic dispersion, which results in high carrier mobility\cite{novoselov_two-dimensional_2005,meric_current_2008} and uniform light absorption of a wide range of wavelengths\cite{mak_measurement_2008,koppens_photodetectors_2014}. Moreover, multilayer graphene shows high electrical and thermal conductance and light absorption with the increase of layer number\cite{zhu_optical_2014,shen_multilayer_2016,negishi_turbostratic_2020}. The screening effect of multilayer graphene also prevents the degradation of the electronic transportation properties caused by the surrounding environment, including charge impurities and substrate roughness\cite{negishi_turbostratic_2020}. However, the energetically favorable AB stacking in multilayer graphene leads to a strong interlayer interaction, which results in a parabolic electronic dispersion\cite{latil_massless_2007}. The intrinsic properties of monolayer graphene, including high carrier mobility, are undermined. Therefore, reducing interlayer interactions allows each layer to exhibit monolayer-graphene-like properties, thereby enhancing the overall material performance of multilayer graphene.

A feasible approach for suppressing interlayer interactions is to increase the interlayer distance. Theoretical analysis has proven that interlayer coupling decreases with the increase of interlayer distance\cite{kinoshita_highly_2017,denner_antichiral_2020}. Experimental observations have shown that the intercalation of small molecules such as $\rm MoCl_5$ and $\rm AlCl_3$ into multilayer graphene reduces the interlayer interaction \cite{kinoshita_highly_2017,lin_coupling_2021}. Nanospacers such as nanodiamonds, carbon nanotubes (CNTs), and nanofibers can also prevent restacking and reduce the interlayer interaction in graphene sponge structures\cite{zhang_enhanced_2012,xu_bulk-scale_2021,xu_stacking_2022}. Compared with other nanospacers, the CNT is another allotrope of carbon with the same $\rm sp^2$ hybridized structure as graphene, which does not introduce impurities and dopants from other elements. Moreover, the graphene and CNT composite system has excellent electronic properties and a high surface-to-volume ratio, which provides widespread potential applications in electronic devices\cite{paulus_charge_2013, liao_enhanced_2019,liu_planar_2015,kholmanov_optical_2015,shang_interfacial_2022}, energy storage\cite{yu_scalable_2014,wang_direct_2020,yu_self-assembled_2010}, chemical catalyst\cite{xu_2d_2020,li_oxygen_2012,zhang_enhanced_2012,shi_facile_2012}, and water treatment\cite{tristan-lopez_large_2013,yin_application_2020}.

The microstructure of the graphene and CNT composite system has a remarkable impact on properties. The addition of CNTs with a diameter of several nanometers can markedly increase the interlayer distance and reduce interlayer interaction\cite{denner_antichiral_2020}. Consequently, each graphene layer is remarkably separated and almost suspended. The benefits of suspended graphene include ultrahigh carrier mobility and ballistic transport\cite{du_approaching_2008,bunch_electromechanical_2007,berciaud_probing_2009,meyer_structure_2007,bolotin_ultrahigh_2008}. However, the single-atom layer structure of graphene results in its low stiffness to bending deformation\cite{lindahl_determination_2012}. The adsorption interaction among graphene layers results in the bending of graphene and a decrease in interlayer distance\cite{fonseca_structure_2019,menon_mechanical_2020,varillas_wrinkle_2021,haraguchi_fabrication_2023,xu_configuration_2023}. Thus, negative factors, including strong interlayer coupling, electron scattering, and a decrease in the surface-to-volume ratio, will appear. Thus, investigating the real effect of CNT insertion on the microstructure of the graphene and CNT composite system is necessary to comprehensively understand and optimize their interaction. In previous experimental studies, the graphene and CNT composite system was usually fabricated through a random mixture of liquid phases, making it difficult to control the layer number and to ensure the CNT insertion between all graphene layers\cite{zhang_enhanced_2012,xu_bulk-scale_2021}. Therefore, a new fabrication approach with enhanced structural controllability must be developed. In addition, molecular dynamics (MD) was employed to investigate the mechanical properties of a multilayer graphene and CNT stacking structure, including elastic constants\cite{menon_mechanical_2020}. The deformation of CNT encapsulated by two graphene layers is also studied theoretically\cite{xu_configuration_2023}. However, the influencing factors that determine the microstructure, such as the size and distribution of CNTs, have not been systematically studied.

In this study, we investigated the multilayer graphene stacking structure with CNT insertion through experiments and simulations. First, we experimentally fabricated such a graphene/CNT stacking structure using a dry transfer method. Highly controlled stacking structures with defined graphene layer numbers are achieved by alternately transferring monolayer graphene and depositing CNTs. Raman spectroscopy confirms a weak interlayer interaction, and atomic force microscopy (AFM) verifies an increase in interlayer distance. Then, MD simulations were conducted to study the microstructure. A systematic investigation of the effect of CNT insertion on the interlayer distance was also conducted, including the CNT distance, diameter, and arrangement. A configuration transition from interlayer suspension to interlayer adsorption with increasing CNT distance was observed. The critical transition distance increases with CNT diameter. Our study provides support for controlling the microstructure of the graphene and CNT composite system and enhancing the material properties.

\section{Methods}
\begin{figure*}[!htbp]
\includegraphics[width=1.0\textwidth]{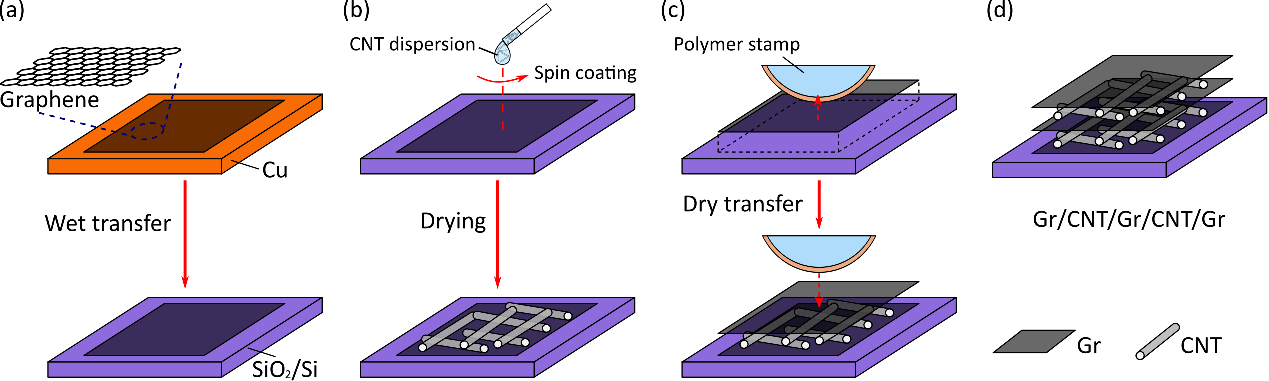}
\centering
\caption{Experimental process of making graphene (Gr) and CNT stacking structure. (a) The CVD graphene is first transferred from Cu foil to Si substrate. (b) The CNT nanospacer is distributed on graphene by spin-coating CNT dispersion. (c) The upper graphene layers are covered by using the dry transfer method. (d) Schematic diagram of the multilayer graphene/CNT stacking structure with three graphene layers (Gr/CNT/Gr/CNT/Gr).}
\label{fig:method}
\end{figure*}

\textbf{Experimental methods.} The graphene/CNT stacking structure was fabricated experimentally by using the dry transfer method (Figure \ref{fig:method}). The monolayer graphene was prepared on Cu foil by chemical vapor deposition (CVD)\cite{brown_polycrystalline_2014}. The graphene was transferred onto a silicon substrate (with 300 nm $\rm SiO_{2}$ oxide layer) using the wet transfer method with a poly(methyl methacrylate) (PMMA) film \cite{li_transfer_2009,suk_transfer_2011}, as shown in Figure \ref{fig:method}a. This step can make graphene on the silicon surface easier to be manipulated in subsequent experiments. Then, the CNT nanospacers were scattered on the graphene surface by spin coating the dispersion of CNT (Figure \ref{fig:method}b). The CNT was synthesized by using the arc-discharge method and dispersed in ethanol with a concentration of 0.05 mg/mL (Meijo Nano Carbon, FH-P). The CNT is single walled with a diameter of $1\text{--}3$ nm according to the manufacturer's information. The sample was placed on a hot plate of 90 $^{\circ}$C to let ethanol evaporate. The areal density of the CNT was increased by repeated spin coating for designated times. The second graphene layer was picked up from the Si substrate, which is prepared by the same wet transfer method and covered onto the CNT nanospacer by using the dry transfer method using polymer stamps\cite{castellanos-gomez_deterministic_2014,pizzocchero_hot_2016,toyoda_pinpoint_2019}, as shown in Figure \ref{fig:method}c. The successful transfer of the second layer can be confirmed by using an optical microscope (Figure S1 in Supporting Information). A multilayer stacking structure can be fabricated by repeating the spin coating of CNT and graphene using a dry transfer process (Figure \ref{fig:method}d). Raman spectroscopy was performed using a $\rm \times100$ objective lens with a laser excitation of 532 nm and a typical spot diameter of $\sim \rm 1 \ \mu m$ (Horiba, LabRAM HR-800). AFM measurements were conducted in the dynamic force mode (Hitachi High-Tech, AFM5100N).

\begin{figure}[!htbp]
\includegraphics[width=0.48\textwidth]{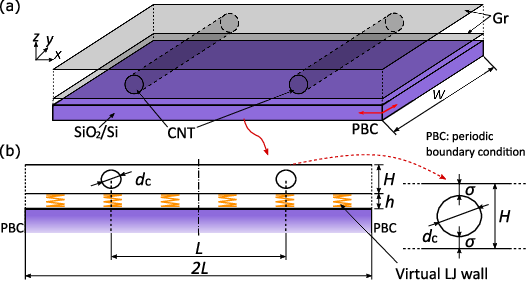}
\centering
\caption{Schematic representation of the simulation system in the three-dimensional view (a) and the front view (b). The system contains two graphene layers and two CNTs aligned parallelly in the $y$ direction. The width of the graphene $W$, equal to the length of CNTs, is fixed. The length of graphene is $2L$ in the $x$ direction. The systems with different CNT distances $L$ and diameters $d_\text{c}$ are simulated. The initial distance between the two graphene layers is $H=d_\text{c}+2\sigma$ as shown in the zoom-up figure. The in-plane $x$ and $y$ directions are set with periodic boundary conditions.
}
\label{fig:MDsystem}
\end{figure}

\textbf{Simulation method.} We conducted MD simulations using LAMMPS program to provide more details on the microstructure of graphene/CNT stacking and the influence of various parameters of CNTs\cite{thompson_lammps_2022}. The simulation model is shown in Figure \ref{fig:MDsystem}. Two parallelly aligned CNTs are encapsulated in two graphene layers (Figure \ref{fig:MDsystem}a). The periodic boundary condition is added in the in-plane $x$ and $y$ directions to simulate the CNT array aligned at equal distances of infinite length. The width of the graphene layers and the length of CNTs are both $W=6.7$ nm in the simulation system. The distance of the CNTs is $L$, and the length of the graphene layers is $2L$ (Figure \ref{fig:MDsystem}b). The graphene interlayer distance of the initial configuration $H=d_\text{c}+2\sigma$, where $d_\text{c}$ is the diameter of the CNTs, and $\sigma$ is the interlayer distance between graphene layers in graphite. Here, we took $d_\text{c}=0.55$ nm, 0.68 nm, 1.08 nm, and 1.35 nm, and $L=10\text{--}70$ nm. The AIREBO potential\cite{stuart_reactive_2000} was used to describe the short-range interaction among carbon atoms. The long-range van der Waals (vdW) interaction among carbon atoms in graphene layers and CNTs is described by using the Lennard-Jones (LJ) potential
 \begin{equation}
E_\text{LJ}=4\epsilon \left( \frac{\sigma ^{12}}{r^{12}}-\frac{\sigma ^6}{r^6} \right) 
\label{eqn:LJ12_6}
\end{equation}
where $r$ is the distance between two atoms. The parameter $\epsilon=0.00284 \ \text{eV}$ is the depth of the potential well, and $\sigma=0.34 \ \text{nm}$ is the zero-energy point and balance distance between graphene layers in graphite\cite{stuart_reactive_2000}. In inducing a vdW interaction between the upper and lower graphene, we set the cutoff radius for long-range interactions larger than the interlayer distance ($r_\text{cutoff} = H + 0.3\sigma$). This process results in increased computational time for simulation. The lower graphene is constrained by a virtual LJ wall to simulate the adsorption effect of the $\rm SiO_{2}$ layer on the Si substrate\cite{aitken_effects_2010,zhang_stiffness-dependent_2015,wang_robust_2019}. The energy of the wall-atom interaction is calculated as follows:
\begin{equation}
E_\text{{sub}}=\frac{\varGamma _0}{\rho _a}\left[ \frac{1}{2}\left( \frac{h_\text{{equ}}}{h} \right) ^9-\frac{3}{2}\left( \frac{h_\text{{equ}}}{h} \right) ^3 \right] 
\end{equation}
where $h$ is the distance between the carbon atom and the virtual LJ wall. The adhesion energy\cite{jiang_measuring_2015,koenig_ultrastrong_2011,gao_measuring_2017} is $\varGamma_0=0.45\ \text{J/m}^2$, and the balance distance between graphene and $\rm SiO_2$ substrate\cite{sonde_dielectric_2009,ishigami_atomic_2007} is $h_\text{{equ}}=0.42$ nm. The areal density of carbon atoms is $\rho _a=38 \ \text{atom/nm}^2$. The system is relaxed in the NVT ensemble at 300 K with a simulation timestep of 1 fs until the energy and configuration of the system reach equilibrium. Thus, the energy-favorable configuration is obtained.

\section{Results and discussion}

\begin{figure*}[!htbp]
\includegraphics[width=0.85\textwidth]{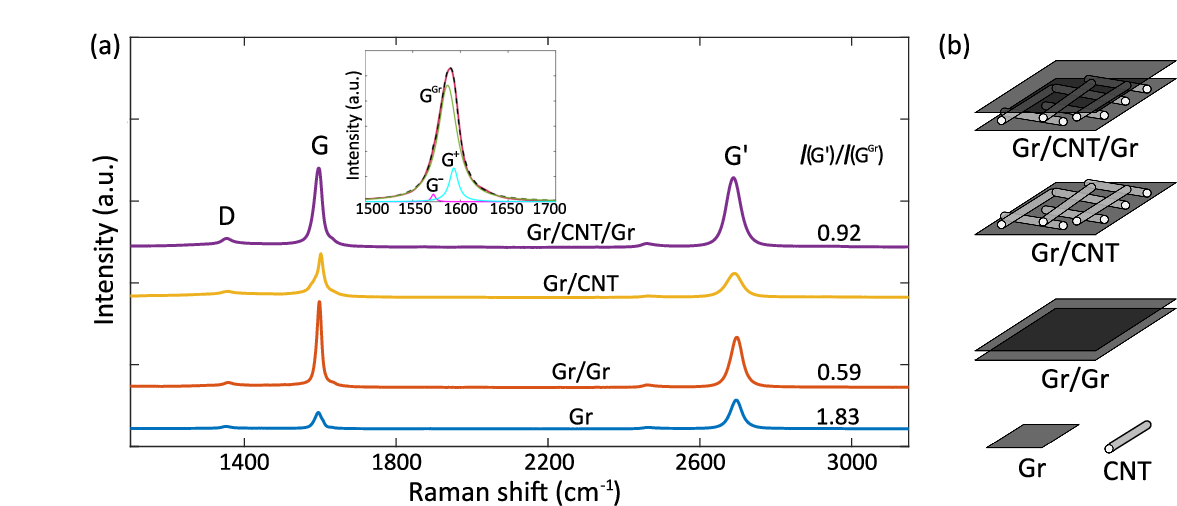}
\centering
\caption{(a) Raman spectrum of Gr (blue), Gr/Gr (red), Gr/CNT (yellow), and Gr/CNT/Gr (purple). The subgraph indicates that the G peak of Gr/CNT/Gr consists of $\rm G^{Gr}$ (at 1580 $\text{cm}^{-1}$) of graphene and $\rm G^-$ (at 1572 $\text{cm}^{-1}$) and $\rm G^+$ (at 1592 $\text{cm}^{-1}$) from single-walled CNTs. The intensity ratio of the $\rm G'$ peak to $\text{G}^\text{Gr}$ peak ($\rm \textit{I}(G')/\textit{I}({G}^{Gr})$) of Gr, Gr/Gr, and Gr/CNT/Gr is indicated. (b) Schematic of the corresponding systems, Gr/Gr, Gr/CNT, and Gr/CNT/Gr, in (a).}
\label{fig:Raman}
\end{figure*}

Raman spectroscopy first validates the weakening of the interlayer interaction of graphene by CNT insertion (Figure \ref{fig:Raman}). Typical Raman spectra of monolayer graphene (Gr) exhibit the D peak ($\sim1350 \ \text{cm}^{-1}$), G peak ($\sim 1580 \ \text{cm}^{-1}$), and $\rm G'$ peak ($\sim2700 \ \text{cm}^{-1}$)\cite{ferrari_raman_2013}. The D peak is affected by crystal defects\cite{ferrari_raman_2013}. The G peak intensity increases with the graphene layer number\cite{koh_reliably_2011,no_layer_2018}. The $\rm G'$ peak relative intensity is sensitive to the graphene layer number and stacking order\cite{ferrari_raman_2006,ni_graphene_2007}. For monolayer graphene, the intensity ratio of the $\rm G'$ peak to the G peak ($\rm \textit{I}(G')/\textit{I}(G)$) is generally $2\text{--}4$\cite{ferrari_raman_2006,ni_graphene_2007,wang_raman_2008,berciaud_probing_2009}. The $\rm \textit{I}({G'})/\textit{I}({G})$ ratio decreases with the increase of the layer number for the AB stacking multilayer graphene due to the strong interlayer interaction and altered band structure with parabolic dispersion. A relatively high $\rm \textit{I}({G'})/\textit{I}({G})$ ratio can be observed in the random stacking multilayer graphene\cite{kim_raman_2012,he_observation_2013,ferrari_raman_2013}. In random stacking graphene, the interlayer interaction is weaker than the AB stacking graphene, and the interlayer distance is slightly larger\cite{kinoshita_highly_2017}. The band structure of the random stacking graphene resembles the monolayer graphene with linear dispersion\cite{latil_massless_2007,denner_antichiral_2020}. The CNT also has the D peak, G peak, and $\rm G'$ peak at similar positions. For single-walled CNTs, the G peak splits into two sub-peaks: $\rm G^-$ (at 1572 $\text{cm}^{-1}$) and $\rm G^+$ (at 1592 $\text{cm}^{-1}$)\cite{paulus_charge_2013}. In our sample, the monolayer graphene shows $\rm \textit{I}({G'})/\textit{I}({G})=1.83$, which is slightly lower than the typical value because of some multilayer islands on the CVD graphene and the doping from substrate impurities and PMMA residue. In the case of bilayer graphene (Gr/Gr), the $\rm \textit{I}({G'})/\textit{I}({G})$ ratio remarkably decreases to 0.59, which is consistent with the previous study\cite{koh_reliably_2011,kim_raman_2012,he_observation_2013}. The $\rm G'$ peak is also lower than the $\rm {G}$ peak in monolayer graphene after adding CNTs (Gr/CNT). The G peak exhibits an asymmetric shape because of the overlapping of the G peak from graphene and split $\rm G^-$ and $\rm G^+$ from CNTs. For the bilayer graphene/CNT stacking structure (Gr/CNT/Gr), the contribution from CNT should be removed when calculating $\rm \textit{I}({G'})/\textit{I}({G})$. We used a three-peak fitting approach to extract the graphene component in the G peak. The fitting result is shown in the subgraph of Figure \ref{fig:Raman}a, and the intensity of the graphene G peak component is referred to as $\rm \textit{I}({G}^{Gr})$. The CNT also shows a $\rm G'$ peak at $\sim2700 \text{cm}^{-1}$. The $\rm G'$ to G ratio is $\sim0.07$ for CNT on $\rm SiO_2/Si$ (Figure S2 in Supporting Information). The proportion of the $\rm G^-$ and $\rm G^+$ peak within the G peak is remarkably smaller than that of the $\rm G^{Gr}$ peak ($\rm \textit{I}({{G}^+})/\textit{I}({{G}^{Gr})}=0.29$). Therefore, we neglected the $\rm G'$ peak induced by the CNT in the total $\rm G'$ peak of the stacking structure. The $\rm \textit{I}({G'})/\textit{I}({{G}^{Gr})}=0.92$ of Gr/CNT/Gr is higher than that of two directly stacked graphene layers (Gr/Gr). This result demonstrates a reduction of interlayer interaction in the Gr/CNT/Gr structure because of the insertion of CNTs.

\begin{figure*}[!htbp]
\includegraphics[width=0.5\textwidth]{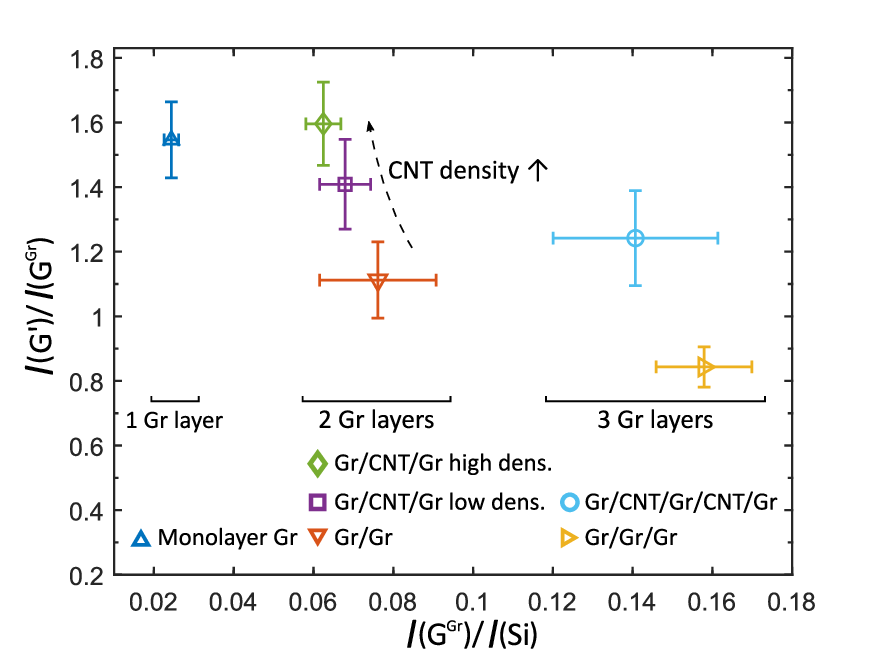}
\centering
\caption{Influence of CNT density and graphene layer number on $\rm \textit{I}({G'})/\textit{I}({{G}^{Gr}})$ and $\rm \textit{I}({{G}^{Gr}})/\textit{I}({Si})$. The average values and standard error bars of $\rm \textit{I}({{G}^{Gr}})/\textit{I}({Si})$ and $\rm \textit{I}({G'})/\textit{I}({{G}^{Gr}})$ are represented in the x-axis and y-axis, respectively. The results include monolayer graphene (blue triangle pointing upwards), Gr/Gr (red triangle pointing downwards), Gr/Gr/Gr (yellow triangle pointing to the right), Gr/CNT/Gr with low density (purple square), Gr/CNT/Gr with high density (green diamond), and Gr/CNT/Gr/CNT/Gr (cyan circle).
}
\label{fig:I2dg}
\end{figure*}

We further investigated the impact of CNT density and layer number on the interlayer interaction by Raman spectroscopy (Figure \ref{fig:I2dg}). The intensity ratio $\rm \textit{I}({{G}^{Gr}})/\textit{I}({Si})$ of the G peak to the substrate silicon peak at $\rm 520 \ \text{cm}^{-1}$ is used to indicate the layer number of graphene, and the interlayer interaction was evaluated by the $\rm \textit{I}({G'})/\textit{I}({{G}^{Gr}})$. Multiple data points were obtained by measuring different sample positions. The average value and standard error were calculated, and the results are plotted in Figure \ref{fig:I2dg}. All point version is shown in Figure S3 in Supporting Information. The increase of $\rm \textit{I}({{G}^{Gr}})/\textit{I}({Si})$ indicates the increase of layer number from Gr to Gr/Gr and trilayer graphene (Gr/Gr/Gr). The ratio does not increase linearly with the number of layers. Previous studies have found that in randomly stacking bilayer graphene\cite{kim_raman_2012,he_observation_2013}, a phonon resonance enhancement and an increase in the G peak intensity are observed when the twist angle between the two layers $\sim12^\circ$. Our multilayer graphene is stacked from polycrystalline CVD graphene with randomly oriented grains, which is also random stacking. Thus, a further increase in $\rm \textit{I}({{G}^{Gr}})/\textit{I}({Si})$ is observed. Comparing the Gr/Gr and Gr/CNT/Gr with different CNT densities, we observed lower values of $\rm \textit{I}({{G}^{Gr}})/\textit{I}({Si})$ with increasing density. Previous studies observed a weaker G peak in suspended graphene compared with graphene on a $\rm SiO_2/Si$ substrate\cite{berciaud_probing_2009,ni_probing_2009}, which is attributed to the effect of substrate. CNTs increase the interlayer distance and make some part of the upper graphene nearly suspended. More CNT nanospacers separate more parts of graphene layers, and the average interlayer distance increases in the measured spots. The partial suspension of graphene contributes to the decrease in $\rm \textit{I}({{G}^{Gr}})/\textit{I}({Si})$. Then, the high CNT density was kept, and the stacking structure of trilayer graphene and CNTs (Gr/CNT/Gr/CNT/Gr) as well as the Gr/Gr/Gr was also fabricated (Figure \ref{fig:I2dg}). The $\rm \textit{I}({G'})/\textit{I}({{G}^{Gr}})$ ratio in the Gr/CNT/Gr/CNT/Gr is higher than that of the Gr/Gr/Gr, demonstrating the continued effectiveness of CNT insertion with increasing layer number.

\begin{figure*}[!htbp]
\includegraphics[width=1.0\textwidth]{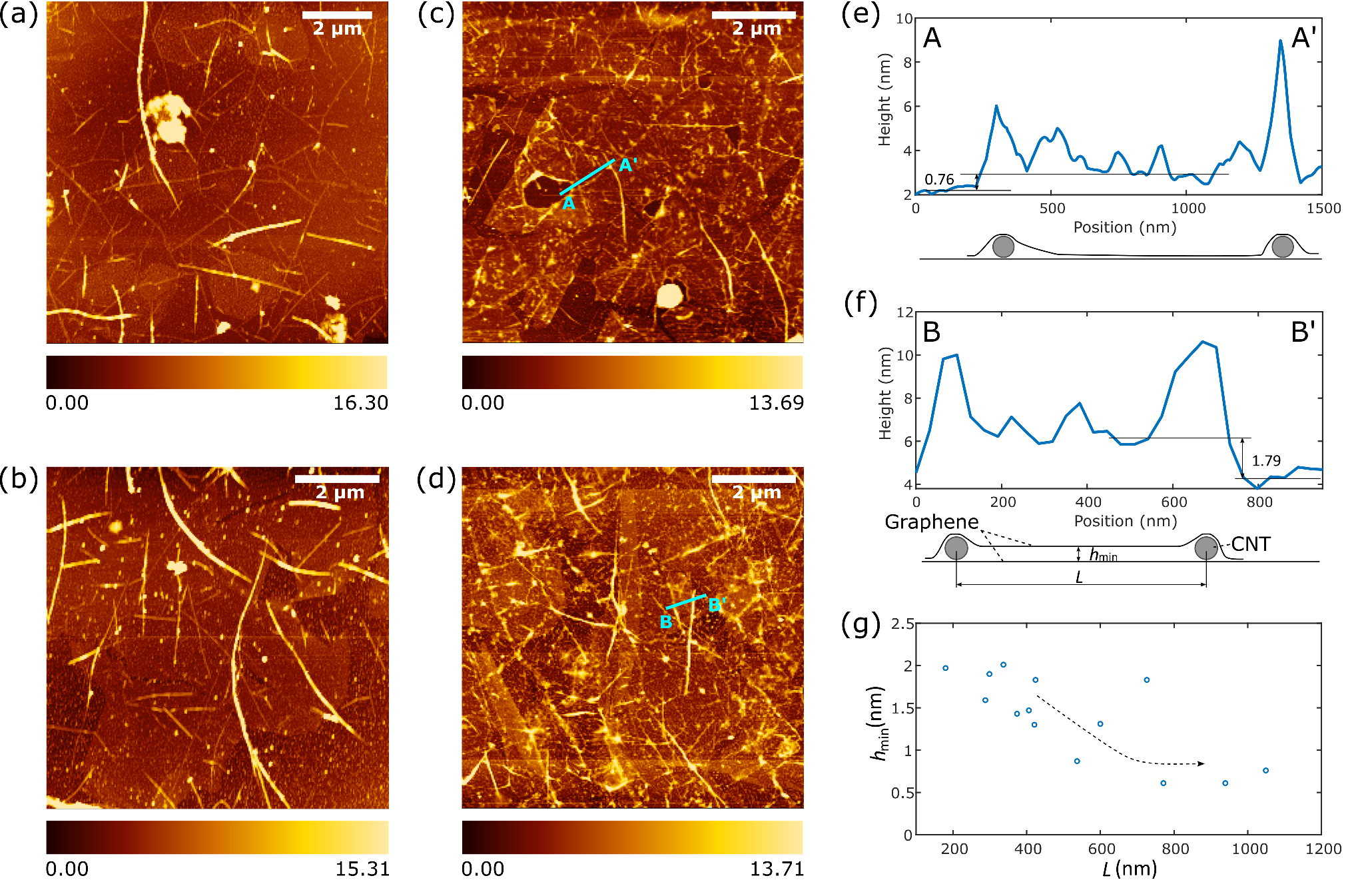}
\centering
\caption{AFM observation of Gr/CNT and Gr/CNT/Gr. (a, b) AFM images of Gr/CNT with low and high densities. (c, d) AFM images of Gr/CNT/Gr with low and high densities. (e) Cross-sectional profile ($\rm A\text{--}A'$) of Gr/CNT/Gr with a longer distance. The profile is selected from (c) with a lower CNT areal density. (f) Cross-sectional profile ($\rm B\text{--}B'$) of Gr/CNT/Gr with a shorter distance. The profile is selected from (d) with a higher CNT areal density. (g) The interlayer distance $h_\text{min}$ between two graphene layers of different CNT distances $L$.}
\label{fig:AFM}
\end{figure*}

The insertion of CNT can reduce the interlayer interaction. The reduction can be attributed to the alteration of the microstructure by increasing the interlayer distance. The AFM measurement was employed to analyze the microstructure of the samples (Figure \ref{fig:AFM}). The AFM images of the CNT distribution with low and high densities on Gr/CNT are shown in Figure \ref{fig:AFM}a,b and Figure S4. The length of the CNTs ranges $1 \text{--} 4 \ \rm \mu m$, with a diameter of $2 \text{--} 8$ nm. The measured diameter is higher than the parameters provided by the manufacturer. The surfactant is added when dissolving the CNTs in ethanol. The surfactant molecules adsorbed on the CNT can increase the measured diameter. The portion without adsorbed surfactant may lead to the attraction among CNTs, forming CNT bundles, which can also cause an increase in the measured diameter. The AFM images of the Gr/CNT/Gr with low and high densities are shown in Figures \ref{fig:AFM}c and d, respectively. By comparing the cross-sectional images (Figures \ref{fig:AFM}e,f), the two CNTs with a larger distance $L$ show a lower thickness ($h_\text{min}$) of the region among CNTs. When the distance is $\sim1000$ nm, the height of the region among CNTs is 0.76 nm. When the distance between the CNTs is $\sim600$ nm, the height of the region is $\sim1.79$ nm, which is remarkably higher than the thickness of two graphene layers. Therefore, the upper layer of graphene is supported by the CNTs to form a partially suspended structure. Figure \ref{fig:AFM}g illustrates the relationship between the distance of CNTs and the height of the upper-layer graphene at multiple cross-sections. As the distance between the CNTs increases, $h_\text{min}$ gradually decreases. A theoretical study has reported that increasing the interlayer distance can weaken interlayer coupling and alter the band structure of multilayer graphene which is comparable with that of a single layer with linear dispersion\cite{denner_antichiral_2020}. CNTs induce the partial suspension of graphene, thereby increasing the interlayer distance. This finding explains the observed reduction of interlayer interactions in Raman spectroscopy.




\begin{figure*}[!htbp]
	\includegraphics[width=1.0\textwidth]{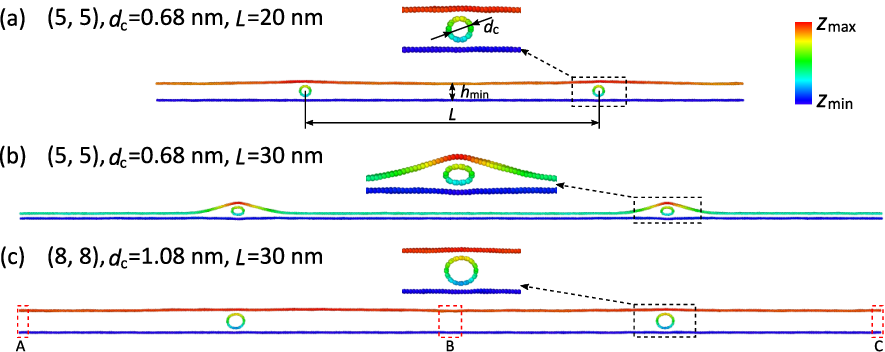}
	\centering
	\caption{Deformed configuration obtained by MD simulation of graphene/CNT stacking structure with different CNT diameters $d_\text{c}$ and distances $L$. (a) The chirality index is (5, 5) and $d_\text{c}=0.68$ nm, $L=20$ nm. (b) The chirality index is (5, 5) and $d_\text{c}=0.68$ nm, $L=30$ nm. (c) The chirality index is (8, 8) and $d_\text{c}=1.08$ nm, $L=30$ nm. The zoom-up figures show the deformation of CNTs. The minimized interlayer distance $h_\text{min}$ is calculated by averaging the distance at A, B, and C in the red dashed frames of (c). The color bar represents the z-coordinate of each atom, where $z_\text{min}$ represented by blue corresponds to the minimum z-coordinate, and $z_\text{max}$ represented by red corresponds to the maximum coordinate.}
	\label{fig:MDdeform}
\end{figure*}

\begin{figure*}[!htbp]
	\includegraphics[width=1.0\textwidth]{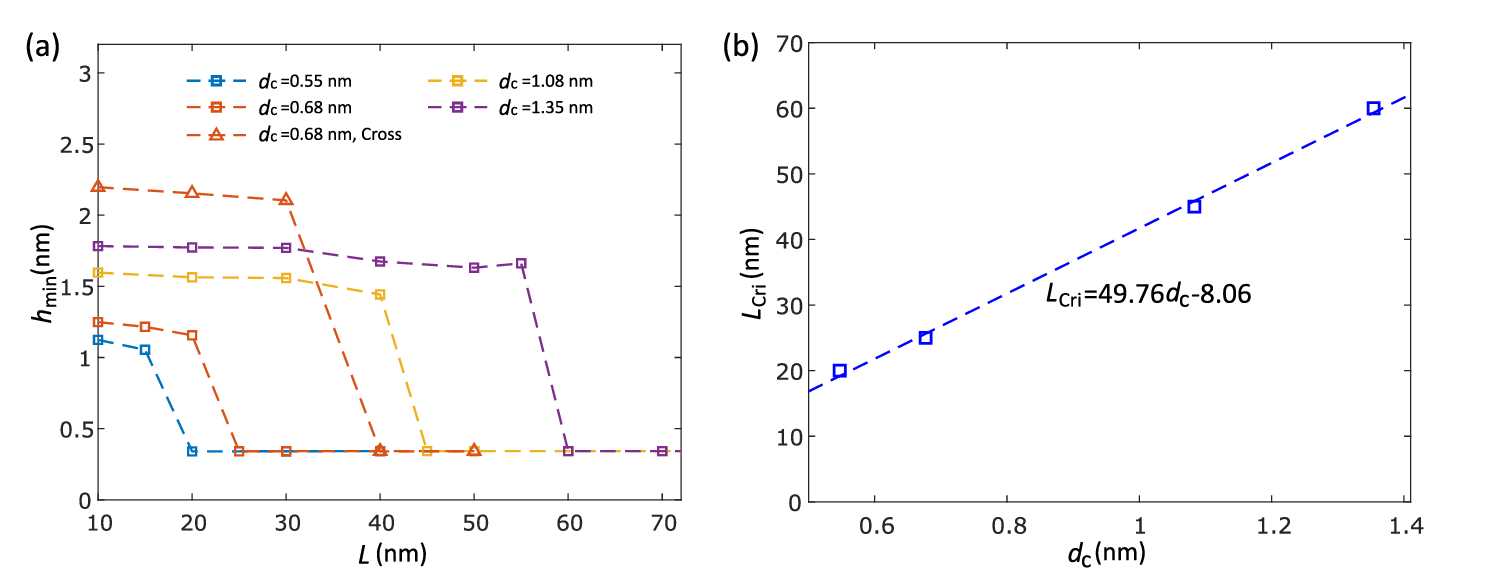}
	\centering
	\caption{(a) Effect of CNT diameter and distance on the minimum interlayer distance. The cases of parallel CNTs are plotted in dashed lines with squares of blue ($d_\text{c}=0.55$ nm), red ($d_\text{c}=0.68$ nm), yellow ($d_\text{c}=1.08$ nm), and purple ($d_\text{c}=1.35$ nm). The cases of crossing CNTs with $d_\text{c}=0.68$ nm are plotted in the red dashed line with triangles. (b) Critical CNT distances ($L_\text{cri}$) when the adsorption of graphene layers functions as the CNT diameter. The blue dashed line indicates the linear fitting result.}
	\label{fig:MDvsModel}
\end{figure*}

We conducted MD simulations of the graphene/CNT stacking structure to investigate the influencing factors of the CNT in detail and to study the microstructure comparatively through AFM observation. The MD simulation can easily control the CNT diameter and area density compared with the experiment. The intuitive microstructure figures can be output at a low cost compared with sample fabrication and AFM scanning. Figure \ref{fig:MDdeform}a--c shows three examples of the MD simulation with different CNT diameters $d_\text{c}$ and CNT distances $L$ ($d_\text{c}=0.68$ nm, $L=20$ nm, $d_\text{c}=0.68$ nm, $L=30$ nm, and $d_\text{c}=1.08$ nm, $L=30$ nm). In the first case (Figure \ref{fig:MDdeform}a), the upper graphene is suspended among the CNTs, but it exhibits a slight curvature after relaxation, which is due to the vdW attraction between the layers. If the distance is increased to $L=30$ nm (Figure \ref{fig:MDdeform}b), then the upper graphene is adsorbed on the lower graphene layer with evident bending deformation near the CNTs. The interlayer distance of the two graphene layers is 0.34 nm, which is close to the spacing of carbon layers in graphite. If the distance $L$ is fixed and $d_\text{c}$ increases to $1.08$ nm (Figure \ref{fig:MDdeform}c), then the upper graphene stays in suspension, which is similar to the first case.

We performed additional simulations by varying the chirality indices, diameters, and distances between the CNTs (Figure \ref{fig:MDvsModel}). The final configurations are classified into two categories: upper graphene suspension and layer adsorption. The minimum interlayer distance $h_\text{min}$ can distinguish between the two configurations. It is calculated by averaging the distance between the graphene layers at the midpoint and two ends (Figure \ref{fig:MDdeform}c). In the case of the same $d_\text{c}$ when the distance $L$ is small (Figure \ref{fig:MDvsModel}a), the $h_\text{min}$ approaches the initial spacing $H$, indicating that the upper graphene layer remains suspended. As the CNT distance gradually increases and surpasses a critical value, the interlayer distance abruptly decreases to 0.34 nm, indicating the occurrence of interlayer adsorption. As the diameter $d_\text{c}$ increases, the interlayer distance $h_\text{min}$ in the suspended graphene configuration also increases. A large CNT diameter also increases the critical CNT distance ($L_\text{cri}$) at which the adsorption configuration appears (Figure \ref{fig:MDvsModel}b). The relation between $d_\text{c}$ and $L_\text{cri}$ is approximately linear. Linear regression is conducted (dashed line in Figure \ref{fig:MDvsModel}b). In the experiments, CNTs are randomly distributed, and crossing CNTs are observed. Therefore, stacking structures with crossing CNTs were also simulated (triangles in Figure \ref{fig:MDvsModel}a). The model is shown in Figure S5. Compared with parallelly aligned CNTs, the crossing CNTs can increase the critical distance for configuration transition from suspension to adsorption.

The transition of configuration from suspension to adsorption configuration implies a change in energy. The potential energy of the simulation system is composed of the internal energy of each graphene layer and CNT described by the AIREBO potential, the vdW energy between the different graphene layers and CNTs described by the LJ potential, and the vdW energy between the lower graphene layer and the $\rm SiO_2$ substrate. The internal energy is influenced by the positional changes of carbon atoms, \textit{i.e.}, deformation of graphene, and CNTs. The vdW energy is determined by the distance between different graphene layers and CNTs (Eq. \ref{eqn:LJ12_6}). Considering that the lower graphene layer remains nearly flat without remarkable deformation, no evident change in its internal energy and vdW energy with $\rm SiO_2$ substrate is observed. The number of atoms in the CNT is smaller than the number of atoms in the graphene layer, and the change in vdW energy induced by CNT deformation is negligible. Thus, the configuration of the system is determined by the competition between the internal energy of the upper graphene and the vdW energy between the two graphene layers. In the suspended configuration, the upper graphene layer has insignificant deformation, resulting in a minor increase in internal energy. However, a higher vdW potential energy contains a larger interlayer distance (Figure S6). For the bilayer adsorption configuration, the vdW energy is minimized with an interlayer distance $h=0.34$ nm. However, the obstructive effect of CNT induces graphene deformation, leading to the bending of angles between C-C bonds and an increase in internal energy. The collapse of CNTs also appears with an ellipse cross-section (see Figure \ref{fig:MDdeform}b), which causes an extra increase in internal energy\cite{xu_configuration_2023}. The adsorption configuration contains higher internal energy and lower vdW energy compared with the suspended configuration.

Experiments and MD simulations have simultaneously discovered that as the distance between CNTs increases, the structure transitions from an upper graphene layer suspension to an interlayer adsorption structure. The trend can qualitatively match, but the quantitative comparison has discrepancies. The critical length $L_\text{cri}$ obtained by the MD simulation is $20\text{--}60$ nm for CNTs with a diameter range of $0.55\text{--}1.35$ nm. Based on the predictions from linear regression analysis, the critical length $L_\text{cri}$ of CNTs with diameters ranging from $2\text{--}8$ nm, which corresponds to the CNT diameters in the experiment, is estimated to be $90\text{--}390$ nm. The experimental observations show graphene suspension with a CNT distance of 700 nm, which is larger than the prediction based on MD simulation. Several factors benefit from the suspension of graphene in the experiment. In ethanol dispersion, surfactant is added to promote CNT dissolution. The surfactant molecules wrap the surface of CNTs by non-covalent interaction, which weakens the interaction between carbon monolayers of graphene and CNTs\cite{vaisman_role_2006}. The distance between CNTs and graphene layers $\sigma$ (Figure \ref{fig:MDsystem}b) is increased because surfactant molecules serve as spacers. Thus, the critical distance $L_\text{cri}$ is increased. Moreover, the CNT is randomly distributed on the surface. The MD simulation indicates that crossing CNTs can increase the critical distance $L_\text{cri}$ (Figure \ref{fig:MDvsModel}a). The CNTs may also aggregate to form CNT bundles where the surfactant is not covered. This effect is similar to the increase in CNT diameter, leading to an increase in $L_\text{cri}$.


\section{Conclusions}
We study the effect of CNT insertion to reduce the interlayer interaction in multilayer graphene. The stacking structure of graphene and CNT is experimentally fabricated. Raman spectroscopy and AFM measurement verify the reduction of interlayer interaction and increase of interlayer distance. In addition, MD simulations are conducted to systematically study the influencing factors. The graphene/CNT stacking structure exhibits two stable configurations: the upper-layer suspension and interlayer adsorption. The distance, diameter, and arrangement of the CNTs determine the selection of the stable structure. Our research contributes not only to the enhanced property of multilayer graphene but also to a comprehensive understanding of the microstructure of the graphene/CNT composite system to facilitate functional enhancement for wider application fields.

\begin{acknowledgement}


This research was supported by JSPS KAKENHI (No. JP15H05866, JP17H05336, JP17H02745, JP19H04545, and JP21H01763), JST SPRING (No. JPMJSP2138), and Murata Research Foundation. The MD simulation of this research work utilized the computing resources of the Institute of Solid State Physics supercomputer center at the University of Tokyo.

\end{acknowledgement}

\begin{suppinfo}

This material is available free of charge via the Internet.

\end{suppinfo}

\bibliography{ref01}

\end{document}


\begin{figure*}[!htbp]
\includegraphics[width=1.0\textwidth]{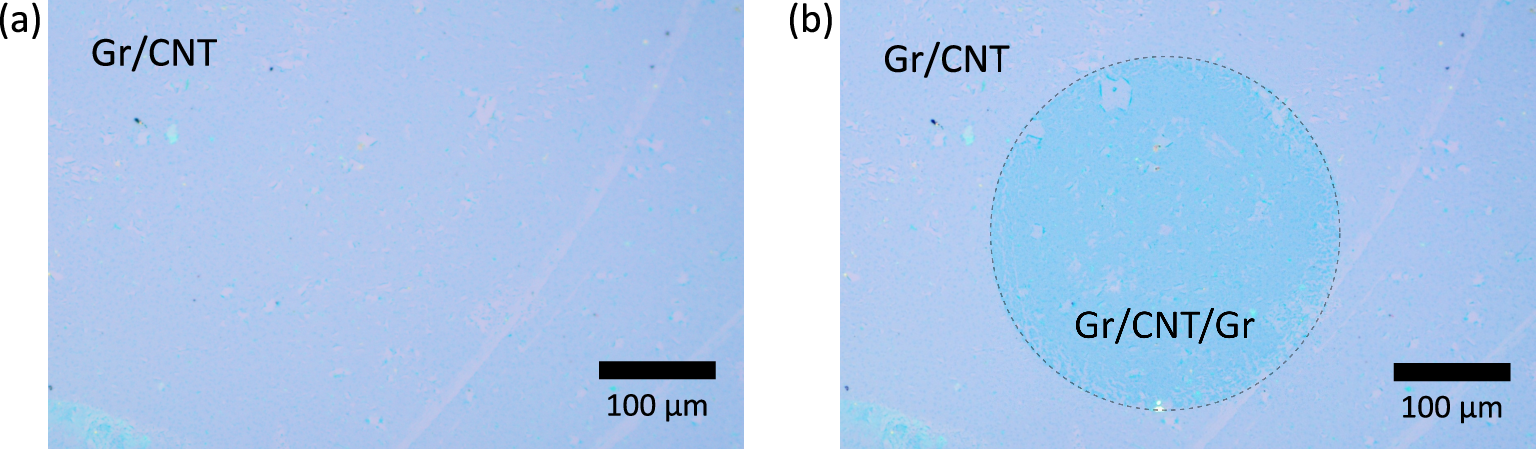}
\centering
\caption{(a) Optical microscope image of Gr/CNT on $\rm SiO_2/Si$ substrate. (b) Optical microscope image of Gr/CNT/Gr after dry transfer. The region in the dashed circle is covered by a second graphene layer.}
\label{fig:supplOM}
\end{figure*}

\begin{figure*}[!htbp]
\includegraphics[width=0.8\textwidth]{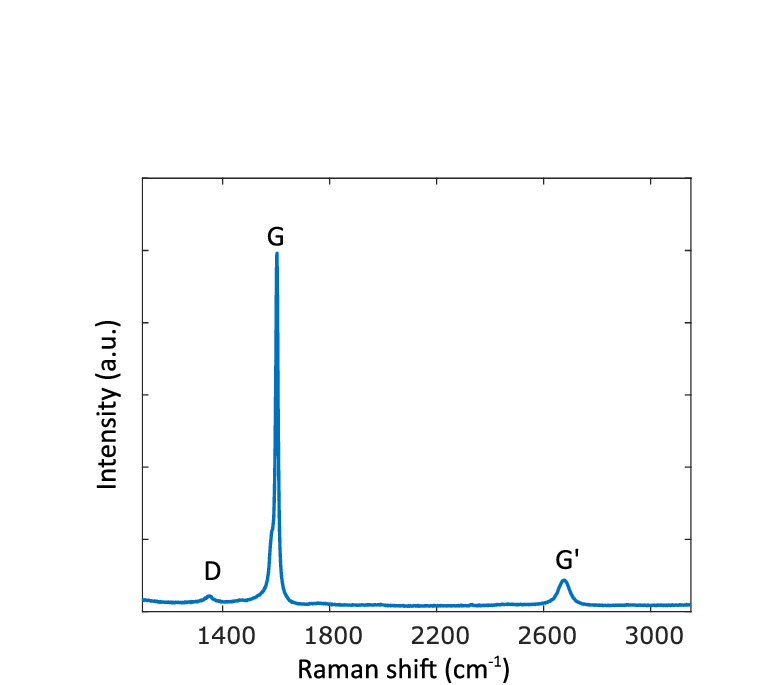}
\centering
\caption{Raman spectrum of CNT. The $\rm \textit{I}({G'})/\textit{I}({G})$ of CNT is 0.07. }
\label{fig:supplRamanCNT}
\end{figure*}

\begin{figure*}[!htbp]
\includegraphics[width=1.0\textwidth]{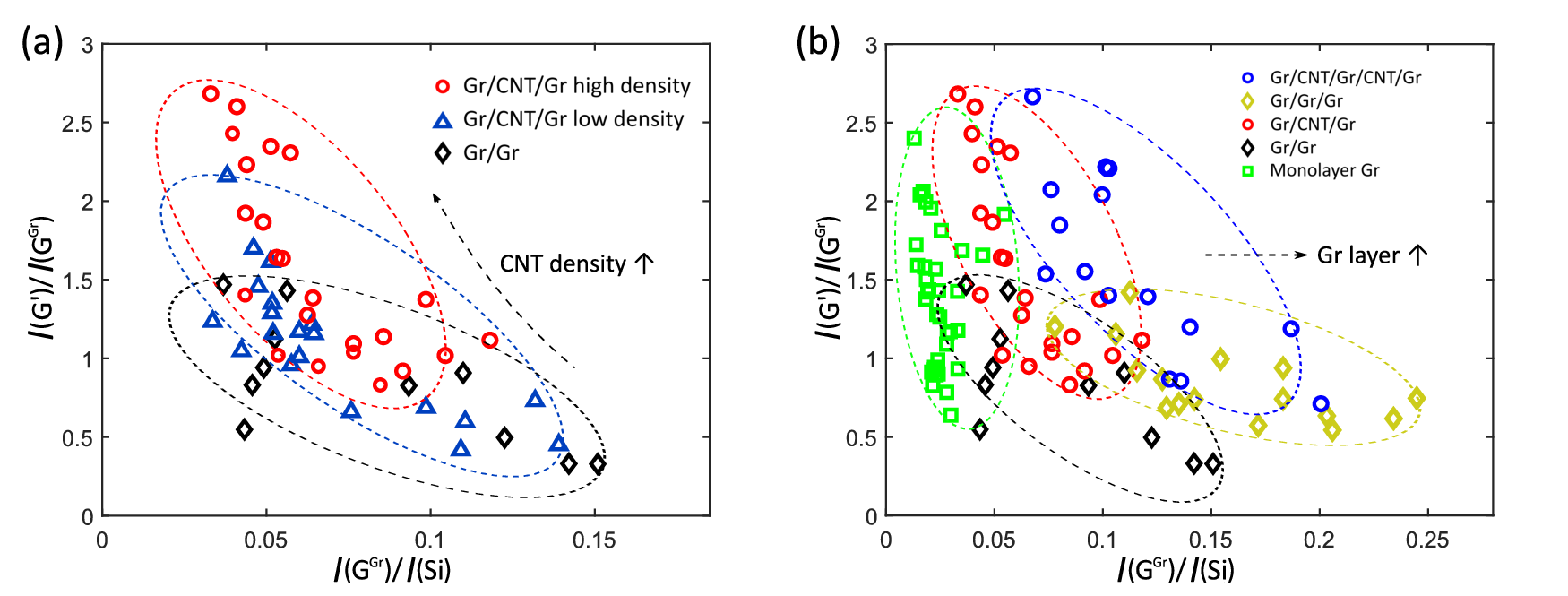}
\centering
\caption{(a) Distribution of $\rm G’/G$ ratio and $\rm G/Si$  ratio for Gr/Gr (black diamond), Gr/CNT/Gr with low density (blue triangle), and Gr/CNT/Gr with high density (red circle). (b) Distribution of $\rm G’/G$ ratio and $\rm G/Si$  ratio for monolayer graphene (green square), Gr/Gr (black diamond), Gr/CNT/Gr (red circle), Gr/Gr/Gr (yellow diamond), and Gr/CNT/Gr/CNT/Gr (blue circle). Only the component of $\text{G}^\text{Gr}$ of graphene for the G peak is calculated in the data points. The data of Gr/Gr and Gr/CNT/Gr in (b) is the high density which matches with those in (a).}
\label{fig:supplI2dgAP}
\end{figure*}

\begin{figure*}[!htbp]
\includegraphics[width=0.8\textwidth]{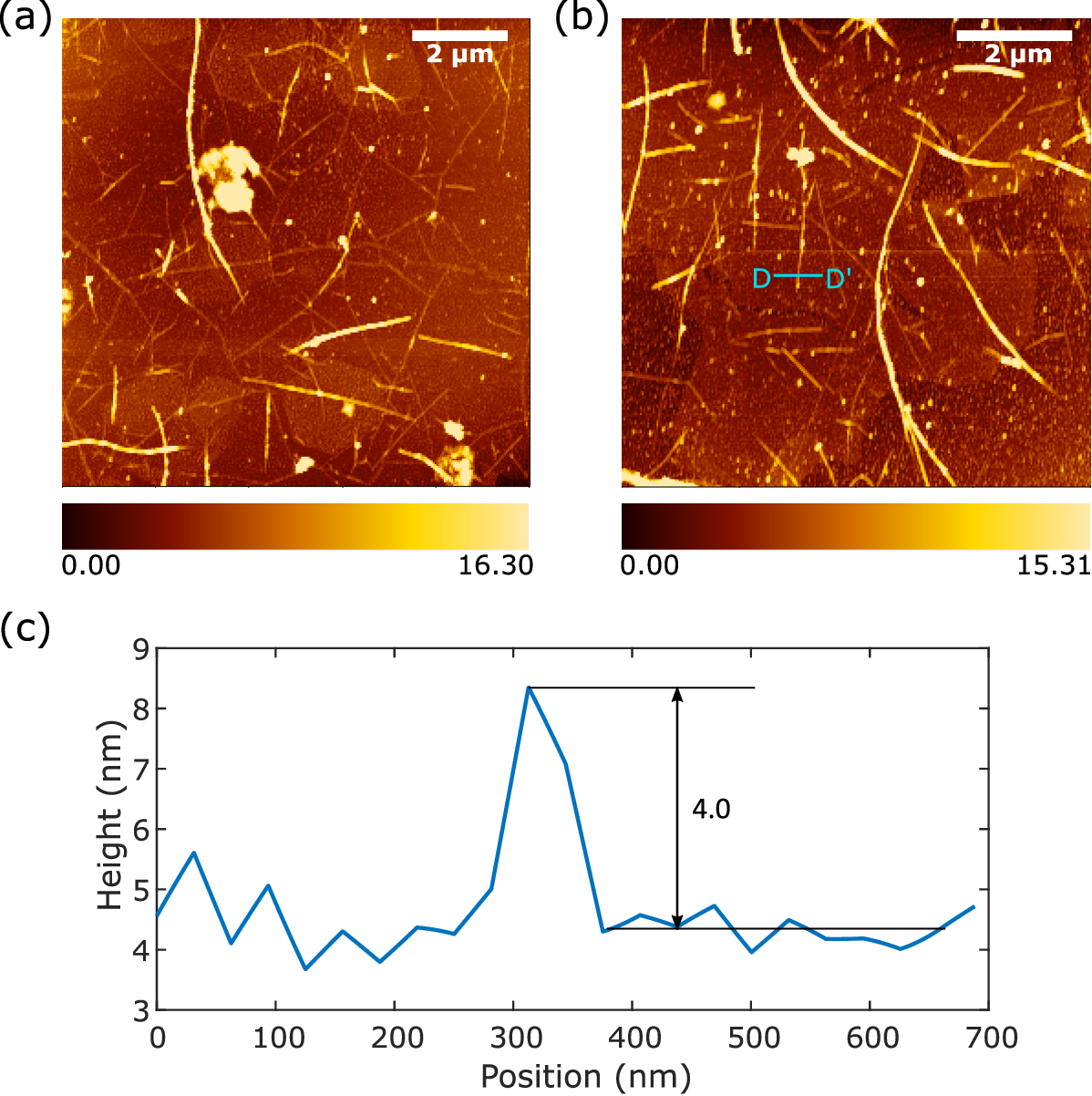}
\centering
\caption{AFM observation of Gr/CNT. (a, b) AFM images of Gr/CNT with low and high densities. (c) Cross-section profile ($\rm D\text{--}D'$) of Gr/CNT from (b). The measured CNT diameter is higher than the data from CNT manufacture because of the surfactant on CNT surface.}
\label{fig:supplAFM}
\end{figure*}


\begin{figure*}[!htbp]
\includegraphics[width=1.0\textwidth]{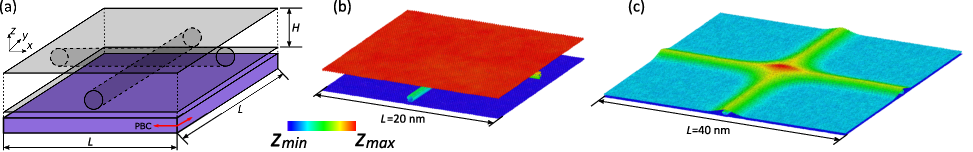}
\centering
\caption{(a) Schematic representation of the simulation system with crossing CNTs. The width and length of the graphene are both $L$. The initial interlayer distance $H=2d_\text{c}+3\sigma$. (b) Deformed configuration of the stacking structure with crossing CNT and $L=20$ nm. (c) Deformed configuration of the stacking structure with crossing CNT and $L=40$ nm. The color bar represents the z-coordinate of each atom, where $z_\text{min}$ represented by blue corresponds to the minimum z-coordinate and $z_\text{max}$ represented by red corresponds to the maximum coordinate.
}
\label{fig:MDsystem}
\end{figure*}

\begin{figure*}[!htbp]
\includegraphics[width=0.8\textwidth]{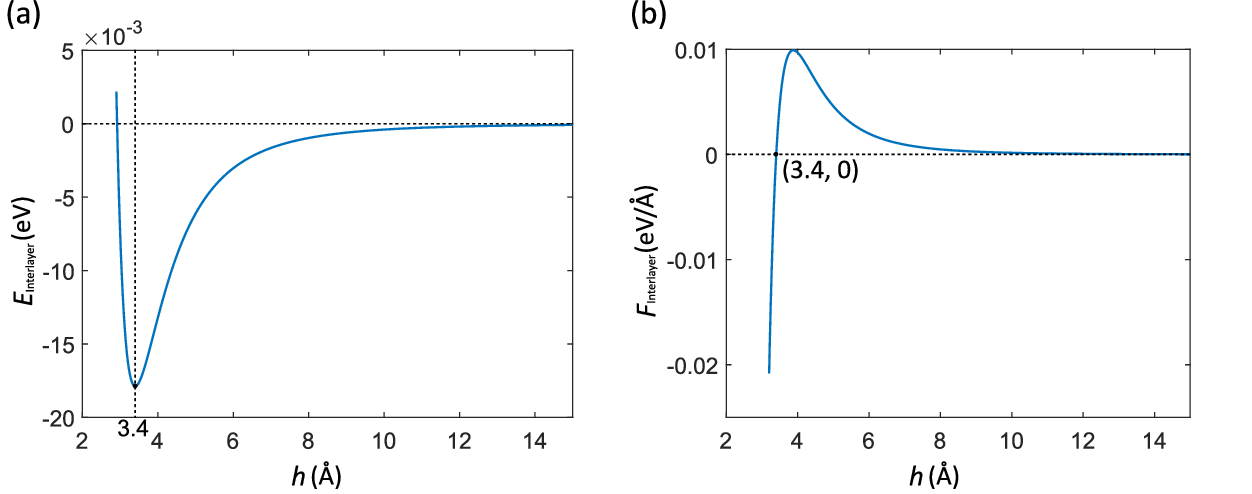}
\centering
\caption{The interlayer energy (a) and attracting force (b) between two graphene layers with distance $h$ of unit area derived from LJ potential.}
\label{fig:supplLJ}
\end{figure*}




